\documentclass{jpp}
\usepackage{graphicx}

\usepackage[utf8]{inputenc}
\usepackage[T1]{fontenc}
\usepackage{amsmath}

\shorttitle{On the evolution of large-amplitude weakly-collisional electron plasma waves}
\shortauthor{A. S. Joglekar and A. G. R. Thomas}

\title{On the evolution of a large-amplitude, weakly-collisional electron plasma wave}

\author{A. S. Joglekar\aff{1, 2, 3}
  \corresp{\email{archis@ergodic.io}},
 \and A. G. R. Thomas\aff{3}}

\affiliation{\aff{1}Ergodic LLC, Seattle, WA 98103 \\
\aff{2}Pasteur Labs, Brooklyn, NY 11205 \\
\aff{3}Department of Nuclear Engineering and Radiological Sciences, 
 University of Michigan - Ann Arbor, 
 Ann Arbor, MI 48109}

\begin{document}

\maketitle

\begin{abstract}
Vlasov-Poisson-Fokker-Planck (VPFP) simulations of large-amplitude electron plasma waves, where the bounce frequency is much larger than the collision frequency, $\omega_B \gg \nu_\text{ee}$, show that the evolution of these waves exhibits three phases; I. A short-lived trapping phase during which collisional effects are minimal. II. A long-lived detrapping phase during which collisional effects are most influential. III. A short-lived Landau damping phase where the effect of collisions becomes minimal again. While the dispersion relation during the trapping and Landau damping phase is well known, the wave behavior during the detrapping phase is not as well understood. The simulations show that during the detrapping phase, the interplay between weak electron-electron collisions and strong wave-electron interactions results in an increasing frequency shift further from the linear root, $\omega_\text{EPW}$. At the conclusion of the detrapping phase, the distribution function is nearly Maxwellian, the frequency shift rapidly diminishes, and the wave damps at a larger rate than the Landau damping rate. Empirical fits to the damping rates, frequency shift enhancement rate, and the lifetime of the plasma waves are provided as functions of collision frequency, wavenumber, and wave amplitude.
\end{abstract}



\section{Introduction}\label{sec:intro}
The dynamics of small-amplitude electron plasma waves (EPW) are described by the complex root, $\omega_L=\omega_\text{EPW} + i \nu_\text{L}$, of the linearized dispersion function $\varepsilon(\omega, k_1; f(\omega)) = 1 - \omega_p^2\int f / (\omega-k_1 v)^2 ~ dv$. Assuming a Maxwellian distribution function, EPW oscillate at a frequency, $\omega_\text{EPW}$, and damp at the Landau damping rate, $\nu_\text{L}$ (\cite{Landau1946, Canosa1973}). In the case of large amplitude EPW, \cite{Bernstein1957} derived a stationary distribution function where the EPW is sustained due to particle trapping in the potential well of the wave. Such a state, often termed a BGK mode, is described by a nearly Maxwell-Boltzmann distribution function and a nonlinear correction due to quasi-linear diffusion (\cite{davidson2012methods}) at the phase velocity of the EPW. \cite{manheimer_formation_1971, dewar_frequency_1972, Morales1972} showed that a BGK mode results in a small negative shift in the wave frequency for a self-consistent solution. In addition to the negative shift in the real part, $\Delta \omega_{NL}$, \cite{oneil_collisionless_1965} shows that the imaginary part, i.e. the damping rate, becomes negligible and the dispersion relation is given by $\omega_{NL} = \omega_\text{EPW} + \Delta \omega_\text{NL} + 0i$.

Previous works assume idealized, collisionless plasmas. Plasmas accessible in the laboratory, e.g. those in fusion and gas discharge experiments, and in Nature, e.g. in the ionosphere and the interstellar medium, are often weakly-collisional such that the ratio of $\nu_\text{ee} / \omega_{p} \sim 10^{-4} - 10^{-10}$ where $\nu_{ee}$ is the electron-electron (e-e) collision frequency (\cite{thorne_modern_2017}).  While it is implicitly known that in collisional systems, BGK modes cannot persist forever and that over time, the wave must exhibit some damping as the particles thermalize, the dynamics of this transition from nonlinear to the linear state are not well understood. 

Using asymptotic analysis, \cite{Zakharov1963} (ZK) determined that during the detrapping phase, the effective damping rate in weakly-collisional, nonlinear EPW can be given by 
$
    \nu_{\text{ZK}} \propto ({\nu_{ee} \omega_\text{EPW}^2}/{\omega_B^3}) \nu_\text{L}
$ 
where $\nu_{ee}$ is the e-e collision frequency, $\omega_{EPW}$ is the linear resonance, $\omega_B$ is the bounce frequency of the trapped particles, $\nu_L$ is the linear Landau damping rate. While the calculation of this relation contains approximations e.g. fluid Landau damping rate, it correctly suggests that waves can have effective damping rates larger than the base e-e collision rate because the rate is inversely proportional to $\omega_B/\omega_p$. Numerical verification of this has been limited to a few simulations using a Particle-In-Cell code where the simulations found qualitative agreement (\cite{valentini_damping_2008}).


This paper provides an end-to-end description of the wave dynamics in weakly collisional, large-amplitude EPW which exhibits three distinct phases. Phase I comprises well-understood collisionless mechanisms of wave excitation, particle trapping, and the establishment of a quasi-linear plateau. The physics in this Phase can be described by a second-order perturbative expansion of the Vlasov-Poisson system with the electric field, $E$, as the small parameter. Phase II is the longest-lived phase during which a quasi-steady state is reached between weak electron-electron collisions and strong wave-electron interactions. Describing Phase II requires a first-order expansion in $\nu_\text{ee}$ as another small parameter in addition to the second-order expansion in $E$. The wave evolution concludes with Phase III when the distribution function has returned to a near-Maxwellian, and the wave energy is rapidly damped via Landau damping. Only the first-order expansion in $E$ is required to describe this Phase accurately. While the behavior in Phases I and III has previously been elucidated, this work describes collisional effects on the wave dispersion relation during Phase II and characterizes the lifetime of Phase II. With these quantities, weakly collisional, large amplitude EPW can be modeled as a piecewise ordinary differential equation. 

The remainder of this paper is organized as follows. Section \ref{sec:sim} describes the simulation configuration. Section \ref{sec:overall} provides an overview of the overall wave dynamics and introduces the three distinct phases of evolution and highlights the enhancement of the nonlinear frequency shift during the detrapping phase. Section \ref{sec:mechanism} describes the physical mechanism for this enhancement of the nonlinear frequency shift. Sections \ref{sec:phase1}, \ref{sec:phase2}, and \ref{sec:phase3} describe each Phase in detail along with empirical fits for the quantities described in table \ref{tab:disp}. Finally, Section \ref{sec:conclusion} summarizes the findings and discusses future work.

\section{Simulation Configuration}\label{sec:sim}
Weakly collisional, large amplitude electron plasma waves are studied using a 1D1V Eulerian Vlasov-Poisson-Fokker-Planck solver (\cite{joglekar_vlapy_2020}) with periodic boundary conditions. The equations that are solved are given by 
\begin{align}
    \frac{\partial f}{\partial t} &+ v \frac{\partial f}{\partial x} - \frac{q}{m} E \frac{\partial f}{\partial v} = \nu_\text{ee} ~ \mathcal{C}(f), \\
    \frac{\partial E}{\partial x} &= \frac{q}{\varepsilon_0} \left(\int_{-\infty}^{\infty} f ~ dv - n_0\right),
\end{align}

While Particle-In-Cell methods are a common tool for studying plasma kinetic behavior, the slow convergence with particle number is a problem for resolving subtle, kinetic phenomena. In these simulations, $N_x = 256, N_v = 2048, L_x = 2\pi/ k\lambda_D, \Delta t = 0.5, \omega_{pe} t_\text{max} = 4000$. Exponential integrators with spectral derivatives in position and velocity space for the different components of the Vlasov equation are used (\cite{thomas_vlasov_2016}). The simulations step through the phase-space integrators with a 6th order integrator in time. Poisson's equation is solved using a spectral solver. The advection-diffusion Fokker-Planck equation (\cite{Dougherty1964}), $\mathcal{C}(f) = \frac{\partial}{\partial v} \left(\bar{v} f + \langle v - \bar{v} \rangle^2 \frac{\partial f}{\partial v}\right)$, is solved using a first-order Euler implicit solver with center differencing of the velocity derivatives. The distribution function is initialized uniformly in space, that is, the density and temperature profiles are uniform. The distribution function is initially Maxwellian.

\begin{figure}
    \centering
    \includegraphics[width=0.5\linewidth]{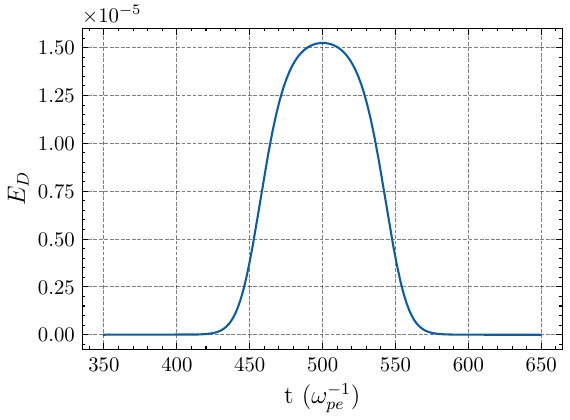}
    \caption{The envelope amplitude of the external forcing term over time. This value is then multiplied by $\sin(k x - \omega_{EPW} t)$ to give the external forcing term on the spacetime grid.}
    \label{fig:driver}
\end{figure}

EPW are driven using a ponderomotive driver as in \cite{joglekar_machine_2023}. Because a Hilbert Transform of the electric field is necessary for determining the time-dependent oscillation frequency, the driver is started $450 \omega_p^{-1}$ into the simulation. This prevents the noise that the Hilbert Transform introduces into the beginning and end of the transformed signal from obfuscating the changing frequency during the drive phase and Phase I. The parameters for the time envelope are $t_c = 500 \omega_p^{-1}, t_w = 100 \omega_p^{-1}, t_R = 25 \omega_p^{-1}, \omega=\omega_\text{EPW}(k\lambda_D)$. This results in a driver profile given in fig. \ref{fig:driver}. The amplitude of this driver profile is multiplied by $\sin(k x - \omega_{EPW} t)$ where $\omega_{EPW}$ is given by the solution to the linearized dispersion relation for small amplitude EPW. 

1600 simulations are performed by scanning over
\begin{align}
    &0.3 \leq k\lambda_D \leq 0.35, \\  
    &0.005 \leq a_0 \leq 0.02, \\
    &10^{-5} \leq \nu_\text{ee} \leq 10^{-3.5},
\end{align}
in 10, 10, and 16 steps, respectively. The parameter ranges were chosen to ensure all three phases are observable within computationally tractable simulation times. For $k \lambda_D < 0.3$, weak Landau damping requires either very long simulations or smaller amplitudes to observe Phase III. For $k \lambda_D > 0.35$, strong damping necessitates larger amplitudes, which would push portions of the parameter space into either purely linear or purely nonlinear regimes.

The following section discusses power-law fits to data collected from the 1600 simulations, and attempts to explain the dynamics that surface in those data.

\section{Overall Wave Dynamics}\label{sec:overall}

\begin{figure}
    \centering
    \includegraphics[width=0.75\linewidth]{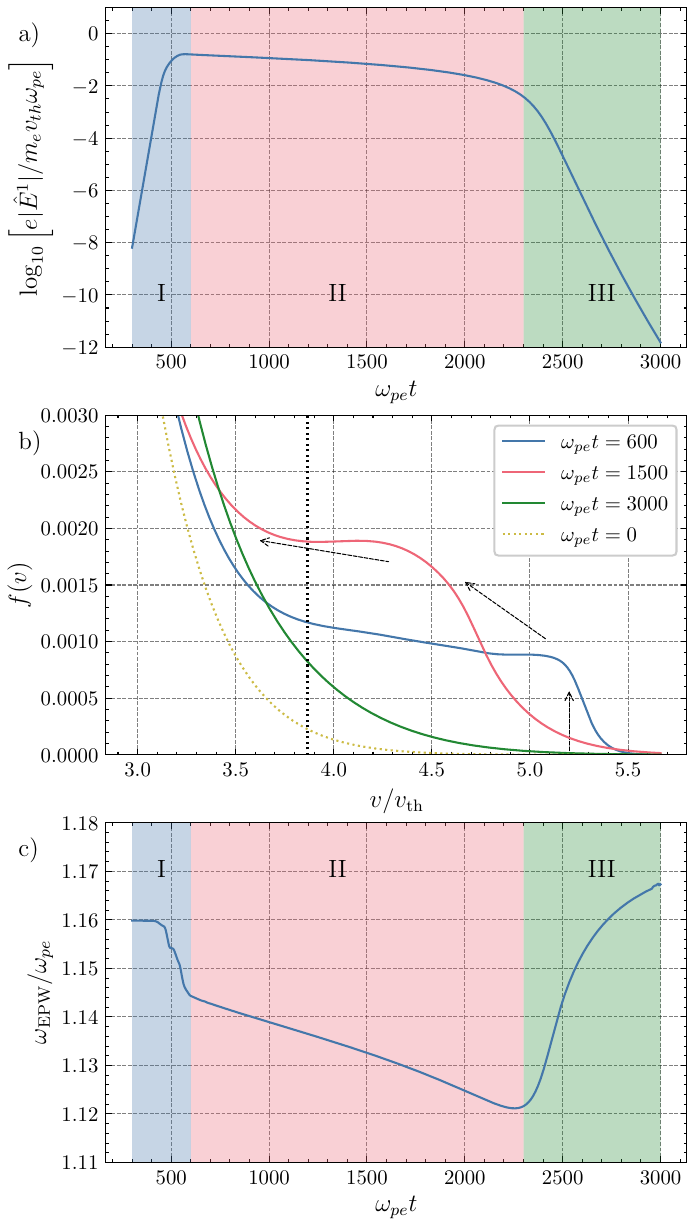}
    \caption{a) Wave amplitude, b) spatially-averaged distribution function and c) wave oscillation frequency are plotted during three distinct phases of wave evolution.}
    \label{fig:phases}
\end{figure}

Figure \ref{fig:phases} shows the three distinct phases of weakly collisional, large-amplitude EPW evolution. Figure \ref{fig:phases}(a) shows the evolution of the wave amplitude. Phase I is very short-lived, as it encompasses the relatively fast dynamics of wave excitation and particle trapping. These dynamics are primarily collisionless, where wave damping is diminished due to quasi-linear diffusion (\cite{Bernstein1957, oneil_collisionless_1965, davidson2012methods}) and there is a corresponding shift in the frequency of oscillation of the wave (\cite{Morales1972, manheimer_formation_1971, dewar_frequency_1972, berger_electron_2013}). Figure \ref{fig:phases}(b) shows the spatially averaged distribution function $|\hat{f}^{0}|(v)$ in the trapping region. During Phase I, it exhibits flattening at the phase velocity of the wave, as expected via quasi-linear diffusion. It also exhibits heating of the particles in a region $\approx \pm 1.5 v_\text{tr}$ from the phase velocity, where the trapping velocity $v_\text{tr} = \sqrt{2E/k}$ and particularly where $v > v_\text{ph} + v_\text{tr}$. This modification of the distribution function in the trapping region results in the well-established nonlinear frequency shift where the EPW frequency downshifts by a few percent as shown in fig. \ref{fig:phases}(c).

During Phase II, fig. \ref{fig:phases}(a) shows that the wave amplitude is weakly damped. Data from simulations indicate that the rate is lower than the Landau damping rate but higher than the collision frequency. In fig. \ref{fig:phases}(b), it is seen that the distribution function has developed a plateau at the phase velocity. The width of the plateau is roughly $v_\text{tr}$. Compared to the distribution function in Phase I, there are more particles in the trapping region and fewer particles in $v > v_\text{ph} + v_\text{tr}$. During this time, the wave frequency clearly decreases, exhibiting a further shift away from the value dictated by linear behavior, $\omega_{EPW} = 1.16$.

After $>\mathcal{O}(100)$ bounce periods, fig. \ref{fig:phases}(a) the damping rate increases rapidly to the order of the Landau damping rate. This indicates the beginning of Phase III. Figure \ref{fig:phases}(b) shows that the trapped particles have thermalized and the distribution function has become nearly Maxwellian. Consequently, fig. \ref{fig:phases}(c) shows that the EPW frequency reverts to that of a small-amplitude, slightly warmer, and strongly damped EPW. 


\begin{table}
    \centering
    \begin{tabular}{c|c|c|l|l}
        Phase & $\mathcal{O}(E^m)$ & $\mathcal{O}(\nu_{ee}^n)$ & $\Re(\omega)$ & $\Im(\omega)$  \\
        \hline
        I  & 2 & 0 & $\omega_\text{EPW} + \Delta \omega_\text{NL}$ & $0$ \\
        II & 2 & 1 & $\omega_\text{EPW} + \Delta \omega_\text{NL}\exp(\gamma_{\Delta \omega} t)$ & $f_\text{NL} ~ \nu_\text{L}$ \\
        III & 1 & 0 & $\omega_\text{EPW}$ & $f_\text{L} ~~ \nu_L$ \\
    \end{tabular}
    \caption{The three distinct phases of nonlinear electron plasma wave evolution are framed in terms of perturbative expansions necessary to describe each phase as well as the resulting dispersion relation of each phase.}
    \label{tab:disp}
\end{table}

The effects on the dispersion relation during the three Phases are summarized in table \ref{tab:disp}, where $\omega_\text{EPW}$ and $\nu_L$ are the real and imaginary part of the solution to the linearized dispersion relation, respectively.  It is found that for $0.3 \leq k \lambda_D \leq 0.35, 0.1 < \omega_B < 0.25$, and $10^{-5} < \nu_{ee} < 3 \times 10^{-4}$, the dynamics can be described by using the set of quantities described in the equations in the table. 

\begin{table*}
    \centering
    \begin{tabular}{c|c|c|c|c|c}
        Quantity & Coefficient & $\alpha(k \lambda_D)$ & $\beta(\omega_B)$ & $\gamma(\nu_{ee})$ & $R^2$ \\
        \hline
        $\Delta \omega_{NL}$ & $93.00 \pm 1.75$ & $5.92 \pm 0.01$ & $0.81\pm 0.00$ & $0.06 \pm 0.00$ & 0.99  \\
        $\gamma_{\Delta \omega}$ & $0.03 \pm 0.00$ & $-0.93 \pm 0.06$  & $-0.88 \pm 0.01$ & $0.80 \pm 0.00$ & 0.97 \\
        $f_{NL}$ & $9.51 \pm 0.87$ & $-1.43 \pm 0.07$  & $-1.48 \pm 0.02$ & $1.07 \pm 0.00$ & 0.99 \\
        $\tau_{II}$ & $0.13 \pm 0.01$ & $-3.82 \pm 0.03$  & $1.23 \pm 0.01$ & $-0.86 \pm 0.00$ & 0.99 \\
        $f_L$ & $1.73 \pm 0.18$ & $-2.25 \pm 0.06$  & $0.78 \pm 0.01$ & $0.15 \pm 0.01$ & 0.92 \\
    \end{tabular}
    \caption{The best fits for the quantities described in table \ref{tab:disp} as power-laws of the form $Q = C (k\lambda_D)^\alpha (\omega_B)^\beta (\nu_{ee})^\gamma$ where $C, \alpha, \beta, \gamma$ are fit parameters. Uncertainties $< 0.005$ are reported as 0.00.}
    \label{tab:fits}
\end{table*}

Table \ref{tab:fits} provides empirical fits for the quantities described in table \ref{tab:disp} as power-laws of the form $Q = C (k\lambda_D)^\alpha (\omega_B)^\beta (\nu_{ee})^\gamma$ where $C, \alpha, \beta, \gamma$ are fit parameters. Each of the fits has an $R^2 > 0.95$ in real and log space indicating that the empirical power laws work well for the parameter space studied here.

The next section focuses on the novel physics in these simulations, the increase in the nonlinear frequency shift due to collisions during Phase II. The subsequent sections describe the dynamics in each Phase in detail and provide empirical fits for the quantities described in table \ref{tab:disp}.

\section{Physical Mechanism for Frequency Shift Enhancement}\label{sec:mechanism}

To determine the reason for the counter-intuitive increase in the frequency shift, the contribution to the dielectric function from the evolution operators in the Vlasov-Poisson-Fokker-Planck (VPFP) system of equations can be calculated. The evolution equation of quasi-linear theory (\cite{davidson2012methods}) for the spatially averaged distribution for a one-component collisional, unmagnetized plasma can be written as

\begin{equation}
    \frac{\partial \hat{f}^0}{\partial t} + \frac{q}{m}\left(\hat{E}^{1\star} \frac{\partial \hat{f}^{1}}{\partial v} + \hat{E}^{1} \frac{\partial \hat{f}^{1\ast}}{\partial v}\right)= \nu_\text{ee} ~\mathcal{C}(\hat{f}^0) \label{eqn:ql0}
\end{equation}

Performing a perturbative expansion assuming $\nu_\text{ee} \ll \omega_B$, that is, the collision frequency is much smaller than the bounce frequency, and using $\hat{f}^0 = \hat{f}^0_0 + \nu_\text{ee} \hat{f}^0_1$ where the subscript indicates the order in $\nu_{ee}$, eq. \ref{eqn:ql0} becomes
\begin{align}
    \frac{\partial \hat{f}^0_0}{\partial t} &+ \frac{q}{m}\left(\hat{E}^{1\star}_0 \frac{\partial \hat{f}^{1}_0}{\partial v} + \hat{E}^{1}_0 \frac{\partial \hat{f}^{1\star}_0}{\partial v}\right) = 0 \label{eqn:ql00}, \\
    \frac{\partial \hat{f}^0_1}{\partial t} &+ \frac{q}{m}\left(\hat{E}^{1\star}_0 \frac{\partial \hat{f}^{1}_1}{\partial v} + \hat{E}^{1}_1 \frac{\partial \hat{f}^{1\star}_0}{\partial v}\right) = \hat{C} \left(\hat{f}^0_0\right) \label{eqn:ql01}.
\end{align}
Equation \ref{eqn:ql00} can be written as the quasilinear diffusion equation that leads to the well-understood collisionless dynamics of the flattening of the distribution function and the nonlinear frequency shift of Phase I. Equation \ref{eqn:ql01} is the weakly-collisional perturbative correction to the collisionless spatially-averaged distribution function $\hat{f}^0_0$. It is $\hat{f}^0_1$ that is responsible for the increase in the frequency shift. That is, the collisional perturbation to the distribution function leads to an enhancement of the nonlinear frequency shift.

This can be formalized by using the dielectric function given by
$
\varepsilon(\omega;k_1) = 1 - \omega_p^2 \int_{-\infty}^{\infty} \frac{\hat{\tilde{f}}^0} {(\omega - k_1 v)^2}dv\;,\label{eqn:disp}
$
where $\hat{\tilde{f}}^0 = \hat{f}^0/n_0$ and $n_0$ is the plasma density, with $\omega_p^2 = q^2n_0/m\varepsilon_0$. If $\hat{\tilde{f}}^0 = \tilde{f}_M$, a Maxwellian distribution, the linear wave dispersion relation has the real frequency $\omega_1$, which is the solution to $\varepsilon(\omega_1;k_1) = 0$. If the dispersion function is expanded about the linear frequency for $\omega = \omega_1 + \delta\omega$, to lowest order,
$
\delta\omega = \omega_p^2\left[\left.\frac{\partial \varepsilon}{\partial\omega}\right|_{\omega=\omega_1}\right]^{-1}\int_{-\infty}^{\infty} \frac{\delta \hat{f}^0}{(\omega_1 - k_1v)^2}dv\;,
$
i.e. the nonlinear change in $\hat{f}^0$ from the coupling of $f_1$ to the field leads to a well known frequency shift (\cite{manheimer_formation_1971, dewar_frequency_1972, Morales1972}).

The time derivative of the frequency shift is
\begin{equation}
\frac{\partial}{\partial t}\delta\omega = \omega_p^2\left[\left.\frac{\partial \varepsilon}{\partial\omega}\right|_{\omega=\omega_1}\right]^{-1}\int_{-\infty}^{\infty} \frac{1}{(\omega_1 - k_1v)^2}\frac{\partial \hat{\tilde{f}}_0}{\partial t}dv\;, \label{eqn:deltaw}
\end{equation}
with ${\partial \hat{\tilde{f}}_0}/{\partial t}$ given by Eqn .~\ref{eqn:ql0} which includes the fast time scale rate of evolution from the normal quasilinear theory (\cite{davidson2012methods}) and the slow collisional relaxation of the distribution function. Expanding eqn. \ref{eqn:deltaw} in $\nu_\text{ee}$ gives
\begin{equation}
    \frac{\partial (\delta \omega_0 + \nu \delta \omega_1)}{\partial t} = \left[\left.\frac{\partial \varepsilon}{\partial\omega}\right|_{\omega=\omega_1}\right]^{-1} \left[\chi(\partial_t \hat{f}^0_{0}) + \nu \chi(\partial_t \hat{f}^0_{1})\right],
\end{equation}
where $\chi(f) = \omega_p^2\int f / (\omega_1 - k_1 v)^2 dv$ is the susceptibility function.

The first term on each side represents the frequency shift described in \cite{Morales1972, dewar_frequency_1972, manheimer_formation_1971}. The second term is the collisional contribution and is given by substituting eqn. \ref{eqn:ql01} for $\partial_t \hat{f}^0_1$ which gives
\begin{align}
    \frac{\partial \delta \omega_1}{\partial t} = \left[\left.\frac{\partial \varepsilon}{\partial\omega}\right|_{\omega_1}\right]^{-1}  \chi \left( \mathcal{V}(\hat{f}^1_{0}, \hat{f}^1_{1})\right) - \chi\left(\mathcal{C}(\hat{f}^0) \right),
\end{align}
where $\mathcal{V}(f_{0}^1, f_{1}^1) = E_{0}^{1*} \frac{\partial f_{1}^1}{\partial v} + E_{1}^{1*} \frac{\partial f_{1}^{1}}{\partial v} + \text{c.c.}$ is the linearized contribution from the Vlasov terms in Eqn. \ref{eqn:ql0}, and $E^1 = E_{0}^1 + \nu E_{1}^1$. By using this relation, the effect on the dielectric function of the Vlasov terms and the Fokker-Planck terms can be isolated and compared. This comparison is performed numerically in fully nonlinear VPFP simulations with a Dougherty collision operator (\cite{Dougherty1964}), given by, $\mathcal{C}(f) = \partial_v (\bar{v} f + \bar{v}_\text{th}^2 \partial_v f)$, where $\bar{v} = \int dv~ v ~f $ and $\bar{v}_{th}^2=\int dv ~ (v-\bar{v})^2 ~ f$. This collision operator conserves density, momentum, and energy and performs a linearized approximation of the phase-space drag and diffusion from small-angle collisions.

\begin{figure}
    \centering
    \includegraphics[width=0.6\linewidth]{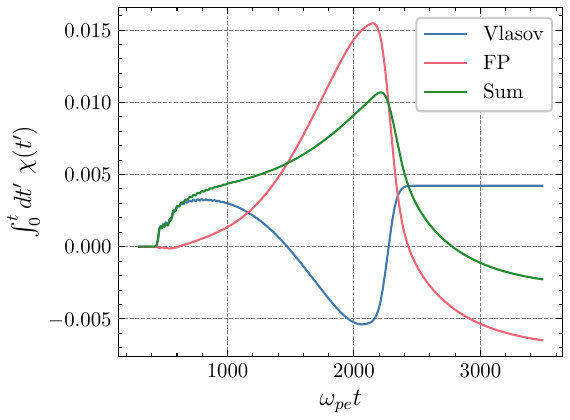}
    \caption{The evolution of the contributions to the susceptibility function}
    \label{fig:deps}
\end{figure}

Figure \ref{fig:deps} compares the change in the dielectric function, $\Delta \epsilon$, from the Vlasov terms to $\Delta \epsilon$ from the collision operator. It shows that during Phase I, the Vlasov contribution dominates as expected from the theory from \cite{Morales1972, dewar_frequency_1972, manheimer_formation_1971}. However, during Phase II, the quasi-steady equilibrium between the Vlasov and Fokker-Planck term is responsible for a net increase in the frequency shift. The Fokker-Planck term always has a positive contribution while the Vlasov contribution counteracts this behavior, but only partially, resulting in a net increase in the frequency shift of the EPW. The subsequent increase can be roughly equal in magnitude to the initial collisionless nonlinear frequency shift from Phase I. This behavior is persistent over all simulations. 

Physically, the collision operator continuously scatters particles at the edges of the trapping region. If the particle trapping is not saturated, there will remain a negative slope at the phase velocity of the wave and wave-particle interactions continue to occur. This quasi-steady competition results in a net redistribution of phase space density that contributes positively to the susceptibility function. The precise mechanism by which this balance produces an enhanced frequency shift, rather than simply restoring linear behavior, remains an open question for future theoretical work.


In the sections that follow, each Phase is described in detail along with empirical fits for the quantities described in table \ref{tab:disp} are provided.


\section{Phase I: Trapping and Plateau Formation}\label{sec:phase1}
Phase I is a short-lived phase that encompasses wave excitation and particle trapping. The modifications to the dispersion relation include the loss of wave damping and a nonlinear frequency shift. In the following sections that describe each Phase in detail, four simulations, each with $k\lambda_D = 0.3, a_0 = 0.1$ and varying collision frequencies are used as examples. These are given in table \ref{table:sims}.

\begin{table}
    \centering
    \begin{tabular}{l | r}
    Simulation & $\nu_{ee} / \omega_{pe}$ \\ \hline
    A & $10^{-5}$ \\
    B & $10^{-4}$ \\
    C & $1.2 \times 10^{-4}$ \\
    D & $3.6 \times 10^{-4}$
    \end{tabular}
    \caption{The simulations used in the remainder of the manuscript have the same $k\lambda_D = 0.3$ and drive strength $a_0$ and varying collision frequencies $\nu_{ee}$}
    \label{table:sims}
\end{table}

\subsection{Phase Space Evolution}
\begin{figure}
    \centering
    \includegraphics[width=0.6\linewidth]{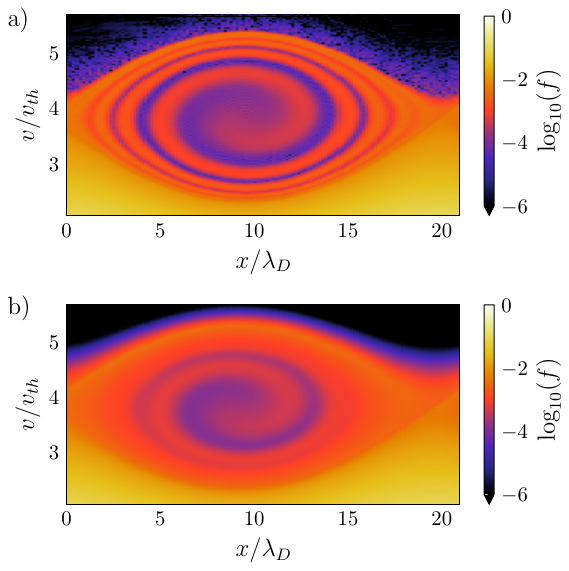}
    \caption{Phase space portrait at $t=650\omega_{pe}^{-1}$, shortly after the drive is turned off, for Simulation A and B.}
    \label{fig:phase1-phasespace}
\end{figure}
During Phase I, particles near the phase velocity of the wave are trapped in the potential well. This creates a phase space vortex near the phase velocity of the wave. 

Phase space snapshots at $\omega_{pe} t = 650$ are shown in fig. \ref{fig:phase1-phasespace}(a) for Simulation A and B. Both snapshots show a similarly sized vortex. This is because the effect of weak collisions and Landau damping has not yet surfaced and the wave amplitudes are nearly the same. However, the snapshot from Simulation B exhibits more smoothing throughout and the features are not as sharp, especially near the edges of the trapping region. 

Figure \ref{fig:phase1-phasespace} also shows that both waves are roughly in phase with one another, as indicated by the fact that the vortex is centered around 10$\lambda_D$ in both cases.



\subsection{Wave Evolution and Nonlinear Frequency Shift}
\begin{figure}
    \centering
    \includegraphics[width=0.6\linewidth]{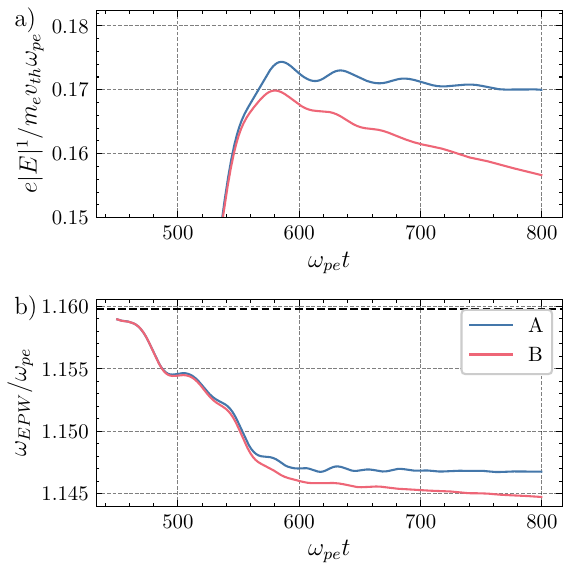}
    \caption{The evolution of the (a) wave amplitude (b) and instantaneous frequency during Phase I for Simulation A and B. The dashed line in (b) indicates the linear frequency.}
    \label{fig:phase-1-amp-wepw}
\end{figure}

The wave evolution early in time in Simulation A and B is plotted in fig. \ref{fig:phase-1-amp-wepw}(a). It shows that the magnitudes of the driven waves in both simulations are within a few percent of one another and only weakly affected by collisions. However, there are two differences that should be highlighted. 

The first is that the effect of the weak damping is already evident in the first $200 \omega_{pe}t$ of the simulation after the drive is turned off around $600 \omega_{pe} t$. The difference in the damping rates is clear in fig. \ref{fig:phase-1-amp-wepw}(a) where the wave amplitude in Simulation B decays faster than that in Simulation A. The instantaneous frequency of the wave is plotted in fig. \ref{fig:phase-1-amp-wepw}(b) and shows similar behavior in that the initial frequency shifts are within a few percent of each other but the decrease in the frequency as the wave amplitude is damped in Simulation B is evident.

Second, the oscillations in the wave amplitude corresponding to particle trapping are suppressed faster in Simulation B. This is likely due to the additional smoothing of the vortex that is evident in fig. \ref{fig:phase1-phasespace}. Similar to the wave amplitude behavior, the bounce oscillations in the wave frequency are suppressed more quickly in Simulation B. 

The nonlinear frequency shift during Phase I is measured by performing a Hilbert Transform of the electric field to extract the instantaneous frequency of oscillation. The frequency shift, $\Delta \omega = \omega - \omega_\text{EPW}$, is measured at the end of Phase I, which is defined as the time when the wave amplitude reaches its first maximum after the drive has been turned off. Fitting the data from all 1600 simulations results in the empirical relation given by
\begin{align}
    \Delta \omega &= 93.00 \pm 1.75 \, (k\lambda_D)^{5.92 \pm 0.01} \, \omega_B^{0.81 \pm 0.00} \, \nu_{ee}^{0.06 \pm 0.00} \label{eq:dwnl}
\end{align}
where $\omega_B = \sqrt{k E}$ is the bounce frequency of the trapped particles. This fit has an R$^2$ value of 0.996 indicating excellent agreement with the data.

The nonlinear frequency shift given by eq. \ref{eq:dwnl} has a strong dependence on $k\lambda_D$ because the frequency shift is a direct result of the rearrangement of the phase space density. The phase space density is a strong function of the phase velocity of the wave which is a strong function of $k\lambda_D$. 

The fact that Phase I comprises collisionless dynamics is also reflected in the effectively negligible dependence on the collision rate in eq. \ref{eq:dwnl}.

\begin{figure}[h]
    \centering
    \includegraphics[width=0.6\linewidth]{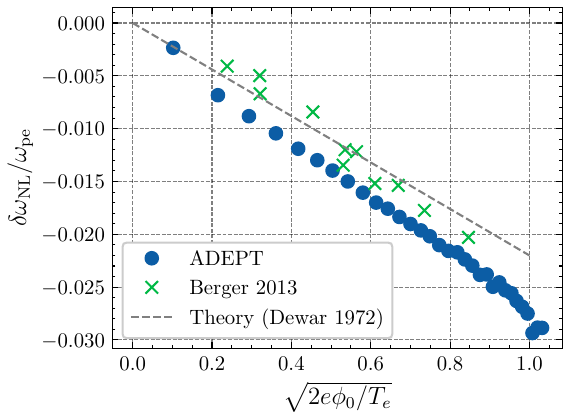}
    \caption{Nonlinear frequency shift vs normalized wave amplitude. Data from \textsc{adept} and \cite{berger_electron_2013} along with the theoretical curve (line) from \cite{dewar_frequency_1972} are compared}
    \label{fig:berger}
\end{figure}

Finally, eq.~\ref{eq:dwnl} has a weaker dependence on the wave amplitude than that reported in \cite{Morales1972, manheimer_formation_1971, dewar_frequency_1972, berger_electron_2013} where $\Delta \omega \propto \omega_B$. While an analytical derivation of this scaling is outside the scope of this work, comparisons to prior simulations and theory are performed. Figure~\ref{fig:berger} shows a direct comparison to the plot in \cite{berger_electron_2013}. The data from \textsc{adept} agrees well with the data from \cite{berger_electron_2013} and both datasets are within a few percent of the theoretical curve from \cite{dewar_frequency_1972}. This figure suggests that \textsc{adept} is performing similarly to the Vlasov code used in \cite{berger_electron_2013} and that the discrepancy between the measured scaling and the theoretical scaling is not due to numerical artifacts but rather due to the approximations used in the derivation of the theory. Further, in \cite{berger_electron_2013}, it is noted that the measured nonlinear frequency shift is sensitive to the specifics of the drive duration. 

It is also important to consider that the linear scaling was derived using asymptotic approximations by \cite{manheimer_formation_1971, dewar_frequency_1972, Morales1972}. Specifically, \cite{Morales1972} stipulate that $v_{tr} v_{ph} \ll v_{th}^2$. In these simulations, $v_{ph} \approx 3.3 v_{th}$ and $v_tr \gtrapprox 0.5$. They also stipulate that their theory requires $v_ph \geq 4 v_{th}$ which leads to $k \lambda_D < 0.3$. In these simulations, $0.3 \leq k \lambda_D \leq 0.35$. These violations of the assumptions used in the derivation of the theory could explain the discrepancy in the scaling with wave amplitude while the close agreement with previous simulations as well as the theoretical curve in fig. \ref{fig:berger} suggests that the simulations are reasonably accurately capturing the nonlinear frequency shift physics.


\subsection{Physical Mechanism for the Nonlinear Frequency Shift}
\begin{figure}
    \centering
    \includegraphics[width=0.6\linewidth]{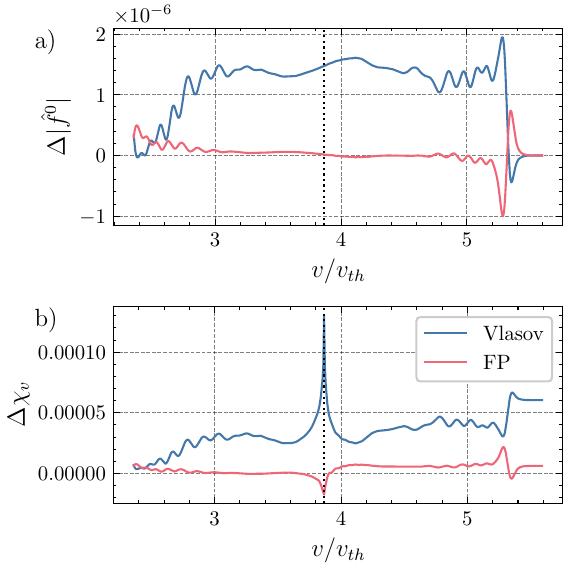}
    \caption{(a) The contribution to $\Delta |\hat{f}^0(v)|$ integrated over the first $\omega_{pe} t = 600$. (b) The corresponding contribution to $\Delta \chi(v)$ which is directly proportional to $\Delta |\hat{f}^0|$}
    \label{fig:phase1-df-dchi}
\end{figure}

 Figure \ref{fig:phase1-df-dchi}(a) distinguishes the change in the spatially averaged distribution function over the first 600 $\omega_{pe}t$ from the Vlasov terms and the Fokker-Planck terms. It shows that the Vlasov terms cause a substantial increase in density in the trapping region while the contribution from the Fokker-Planck terms is negligible. 
 
 Figure \ref{fig:phase1-df-dchi}(b) shows the corresponding contribution to the susceptibility function, $\Delta \chi(v) = \omega_{pe}^2 \int_0^v \partial_v' \Delta |\hat{f}^0(v')| / (\omega - k v') dv'$, which is directly proportional to the frequency shift. The different contributions are calculated by summing the change in the distribution function from each of the terms in the simulations. It shows that the Vlasov terms are responsible for the nonlinear frequency shift while the Fokker-Planck terms have a negligible effect. This confirms that the nonlinear frequency shift during Phase I is a collisionless effect arising from wave-particle interactions, as expected from \cite{Morales1972, manheimer_formation_1971, dewar_frequency_1972, berger_electron_2013}.





\section{Phase II: Quasi-Steady Collisional Phase}\label{sec:phase2}
Phase II is the longest-lived phase during which a quasi-steady state between weak electron-electron collisions and strong wave-electron interactions governs the dynamics. Figure \ref{fig:phases}(a) and (c) suggests that the wave evolution during this phase is characterized by three quantities; the effective damping rate of the wave, the growth rate of the nonlinear frequency shift, and the lifetime of Phase II. Each of these quantities is discussed in the following subsections. First, the phase space evolution during this phase is described.

\subsection{Phase Space Evolution}
\begin{figure}
    \centering
    \includegraphics[width=0.6\linewidth]{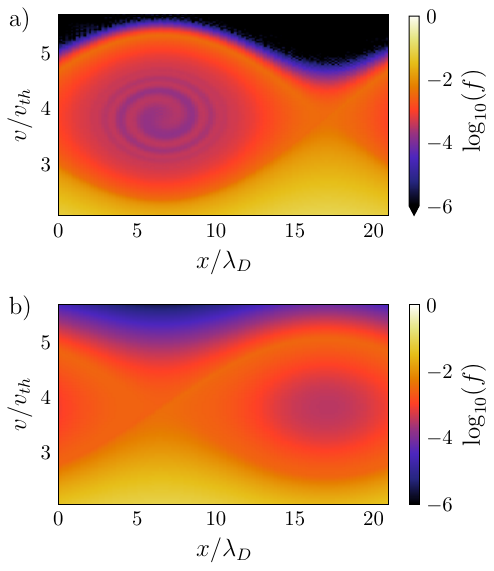}
    \caption{Phase space portrait at $t=1400\omega_{pe}^{-1}$, long after the drive is turned off, for two different collision frequencies.}
    \label{fig:phase-2-phasespace}
\end{figure}

Figure \ref{fig:phase-2-phasespace}(a) and (b) shows phase space snapshots during Phase II for Simulation A and Simulation B, respectively, at $\omega_{pe}t = 1400$. 

The snapshots show similarly sized vortices persist in both simulations despite the wave amplitude in Simulation B being roughly half that of Simulation A. 

Simulation B exhibits substantial phase space diffusion to the extent that the trapped particle orbits are not discernible as they are in Simulation A. This is because the higher collision frequency in Simulation B leads to more rapid phase space diffusion in the deeply trapped region where the gradients in velocity space are large. 

It is also evident from fig. \ref{fig:phase-2-phasespace} that there is a phase shift between the two simulations. This is because the wave frequency has experienced a greater downshift in Simulation B due to the effect highlighted in the previous section.

\subsection{Wave Amplitude and Frequency Evolution}
\begin{figure}
    \centering
    \includegraphics[width=0.6\linewidth]{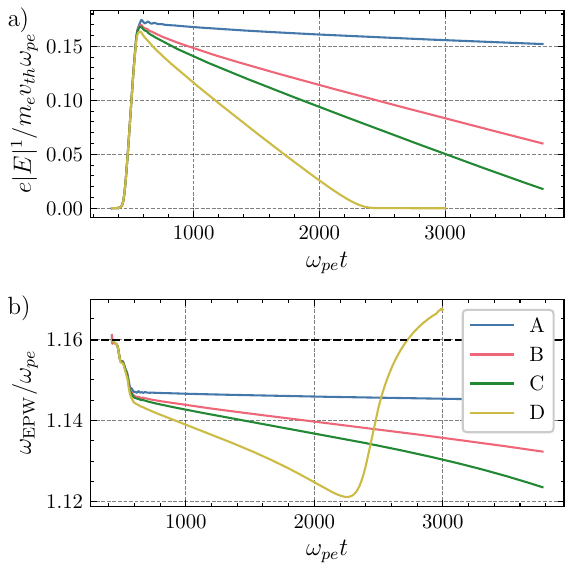}
    \caption{Evolution of the (a) wave amplitude and (b) wave frequency for 4 different collision frequencies.}
    \label{fig:phase-2-amp-wepw}
\end{figure}

Figure \ref{fig:phase-2-amp-wepw}(a) shows the evolution of the electric field amplitude over a longer time scale for all four simulations. The trend here is as expected; with increasing collision frequency, the effective damping rate increases. Bounce oscillations have been suppressed over this time scale. While not shown in fig. \ref{fig:phase-2-amp-wepw}(a), fig. \ref{fig:phases}(a) shows that the wave amplitude in Simulation D decreases rapidly after $\omega_{pe}t \approx 2300$ indicating the beginning of Phase III.

Figure \ref{fig:phase-2-amp-wepw}(b) shows the corresponding evolution of the instantaneous frequency of the wave. It shows that the frequency can downshift substantially during Phase II. Interestingly, the rate of downshift increases with increasing collision frequency. Similar to the wave amplitude behavior, fig. \ref{fig:phases}(c) shows that the frequency in Simulation D begins to rapidly increase after $\omega_{pe}t \approx 2300$ indicating the beginning of Phase III.

\subsection{Physical Mechanism for Frequency Shift Enhancement}

\begin{figure}
    \centering
    \includegraphics[width=0.6\linewidth]{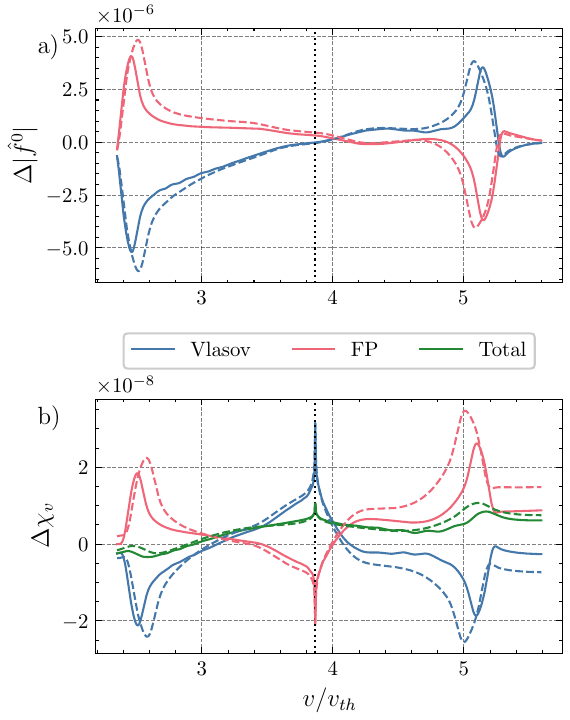}
    \caption{(a) The change in phase space density $\Delta |\hat{f}^0(v)|$ over Phase II from the different terms. (b) The corresponding contributions to the susceptibility function $\Delta \chi(v)$ and the sum. Solid - Sim A, Dashed - Sim B.}
    \label{fig:phase2-df-dchi}
\end{figure}

Figure \ref{fig:phase2-df-dchi}(a) distinguishes the change in the spatially averaged distribution function over the first 600 $\omega_{pe}t$ from the Vlasov terms and the Fokker-Planck terms. It shows that the Vlasov terms redistribute the phase space density from slower velocities to faster velocities while the Fokker-Planck terms do the opposite. The net result, not shown here due to space constraints, is a slight increase in phase space density in the trapping region above the phase velocity. 
 
Figure \ref{fig:phase2-df-dchi}(b) shows the corresponding contribution to the susceptibility function. It shows that the Vlasov terms, in blue, are responsible for a net \emph{decrease} in the frequency shift while the Fokker-Planck terms, in red, are responsible for a net \emph{increase} in the frequency shift. The net result, in green, is an increase in the frequency shift during Phase II. 

The dashed lines in fig.~\ref{fig:phase2-df-dchi} are the results from Simulation B. It shows similar behavior to Simulation A, but the magnitudes of the contributions are larger due to the higher collision frequency. A larger increase in phase space density in the trapping region results in a larger increase in the frequency shift.

\subsection{Summary of Phase II Dynamics}
In Phase II, the growth rate of the nonlinear frequency shift, the effective damping rate of the wave, and the lifetime of Phase II, are given by
\begin{align}
    \gamma_{\Delta \omega} &= 0.03 \pm 0.00 \, \frac{\nu_{ee}^{0.80 \pm 0.00}}{(k\lambda_D)^{0.93 \pm 0.06} \, \omega_B^{0.88 \pm 0.01}}, \label{eq:dwnlrate} \\
    f_\text{NL} &= 9.51 \pm 0.87 \, \frac{\nu_{ee}^{1.07 \pm 0.00}}{\, (k\lambda_D)^{1.43 \pm 0.07} \, \omega_B^{1.48 \pm 0.02} }, \label{eq:damprate} \\
    \tau_\text{ II} &= 0.13 \pm 0.01 \, \frac{\omega_B^{1.23 \pm 0.01}}{\, (k\lambda_D)^{3.82 \pm 0.03}  \nu_{ee}^{0.86 \pm 0.00}}, \label{eq:lifetime}
\end{align}
respectively.  Equations \ref{eq:dwnlrate} and \ref{eq:damprate} vanish as $\nu_\text{ee} \rightarrow 0$ and the collisionless dynamics of Phase I are recovered. Collisionless behavior prescribes a persistent wave, with a frequency shift $\Delta \omega_\text{NL}$ due to a trapped particle distribution. 

The damping rate in eqn. \ref{eq:damprate} has a weaker scaling than predicted in \cite{Zakharov1963}. Their derivation results in a $\omega_B^{-3}$ scaling, while the best fit finds a weaker scaling of $\omega_B^{-1.48}$. Due to this, the ZK damping rate significantly underestimates the observed damping rate for large $\omega_B$. This could be because ZK uses a perturbative approach that assumes strong trapping such that $\omega_B \gg \nu_{ee} v_{th}^2/v_{tr}^2$, which may not hold for weaker wave amplitudes. Similar to the nonlinear frequency shift, it might be that $\omega_B^{-3}$ scaling would be more accurate for much larger wave amplitudes, and at smaller wavenumbers. 

Simulations show that the EPW persists in Phase II for $>\mathcal{O}(10)$ trapping oscillations. The lifetime of the quasi-steady state during Phase II can also be characterized and an empirical fit is provided in eqn. \ref{eq:lifetime}.

\section{Phase III: Return to Landau Damping}\label{sec:phase3}
Phase III begins when the distribution function has returned to a nearly Maxwellian state and the wave rapidly damps via Landau damping. As shown in fig. \ref{fig:phases}(a), the damping rate increases rapidly to approximately the Landau damping rate (or slightly higher), and fig. \ref{fig:phases}(c) shows that the oscillation frequency rapidly returns to a value near the initial $\omega_\text{EPW}$.

\begin{figure}
    \centering
    \includegraphics[width=0.6\linewidth]{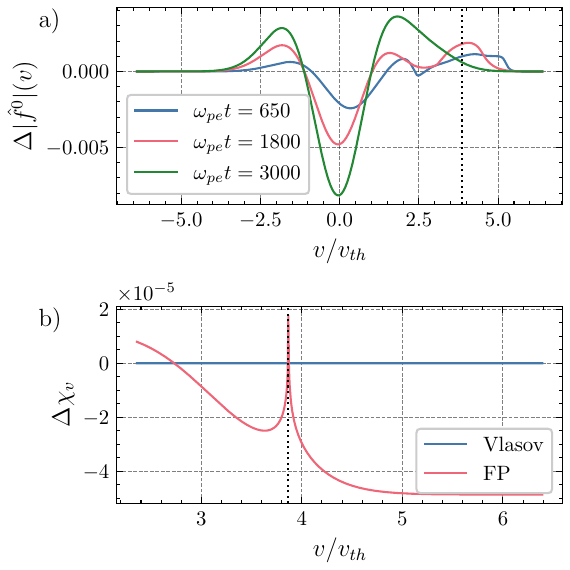}
    \caption{(a) $\Delta|\hat{f}^0(v)|$ accumulated through Phase I, Phase II, and Phase III. (b) The contribution to $\Delta \chi(v)$ in Phase III from each term.}
    \label{fig:phase3-df-dchi}
\end{figure}

\subsection{Distribution Function Thermalization}

Figure \ref{fig:phase3-df-dchi}(a) shows the total change in the spatially averaged distribution function over all three phases for Simulation C. During Phase I and II, there is an increase in density near the trapping velocity. During Phase III, this density increase has been erased and the additional phase space density has been redistributed to lower velocities as the distribution function returns to a Maxwellian.

Figure \ref{fig:phase3-df-dchi}(b) shows the contribution to the susceptibility function from each term during Phase III. It shows that the Fokker-Planck term is responsible for returning the distribution function to a Maxwellian, while the effect of the Vlasov term is negligible.

\subsection{Wave Properties and Enhanced Damping}

Finally, the Phase III dynamics are given by
\begin{align}
    f_\text{L} &= 1.73 \pm 0.18 \, \frac{\omega_B^{0.78 \pm 0.01} \, \nu_{ee}^{0.15 \pm 0.01}}{(k\lambda_D)^{2.25 \pm 0.06}}, \label{eq:fL}
\end{align}
where $f_L$ gives the damping rate as a multiple of the Landau damping rate in Phase III. It is seen that with more energy absorbed by the particles during Phase II (i.e. larger $\omega_B$), the damping rate during Phase III is larger because the final Maxwellian is slightly warmer.

During Phase III, $\Delta \epsilon$ relaxes to a value smaller than the value at $t=0$, suggesting a small frequency \emph{upshift}. This upshift is likely due to the heating of the distribution function from absorbing the EPW energy. The return to a Maxwellian distribution is gradual during Phase III as the Fokker-Planck contribution asymptotically approaches the necessary value for the slightly warmer EPW.

\section{Conclusion}\label{sec:conclusion}

This work presents a comprehensive kinetic study of large-amplitude electron plasma wave evolution in weakly collisional plasmas using 1600 Vlasov-Poisson-Fokker-Planck simulations spanning a broad parameter space in wavenumber, wave amplitude, and collision frequency. The simulations reveal that the wave dynamics naturally partition into three distinct phases, each governed by different physical mechanisms and amenable to different perturbative descriptions.

Phase I encompasses the well-understood collisionless dynamics of wave excitation, particle trapping, and quasi-linear plateau formation. During this short-lived phase, the distribution function flattens at the phase velocity, wave damping is suppressed, and a nonlinear frequency downshift emerges consistent with previous theoretical predictions, though with a weaker amplitude scaling ($\Delta \omega \propto \omega_B^{0.82}$) than the linear dependence predicted by asymptotic theory.

The central finding of this work concerns Phase II, during which a quasi-steady equilibrium develops between weak electron-electron collisions and strong wave-electron interactions. Counter-intuitively, collisions are found to enhance the nonlinear frequency shift rather than simply restore linear behavior. This effect arises because the Fokker-Planck collision operator contributes positively to the susceptibility function, while the Vlasov terms only partially counteract this contribution. The frequency shift can approximately double during Phase II relative to its Phase I value. Empirical scaling relations for the initial frequency shift (eq. \ref{eq:dwnl}), frequency shift growth rate (eq. \ref{eq:dwnlrate}), effective damping rate (eq.~\ref{eq:damprate}), and phase lifetime (eq. \ref{eq:lifetime}) are provided, enabling the construction of reduced models for weakly collisional EPW dynamics.

A comparison of the damping rate in Phase II with Zakharov-Karpman theory reveals that while the linear scaling with collision frequency is confirmed, the theory underpredicts the damping rate by approximately an order of magnitude and predicts a stronger inverse dependence on bounce frequency ($\omega_B^{-3}$) than is observed ($\omega_B^{-1.48}$).

Phase III marks the return to linear behavior as collisions restore a near-Maxwellian distribution and the wave undergoes rapid Landau damping. The transition occurs after a time, $\tau_{II}$ given by eq. \ref{eq:lifetime}, and the damping rate during this phase is characterized by eq. \ref{eq:fL}


Future work should extend this analysis to explore the role of ion dynamics, and investigate the interaction of multiple EPW modes in the weakly collisional regime. Additionally, developing analytic theory that captures the enhanced frequency shift mechanism identified here would provide deeper physical insight into the quasi-steady collisional dynamics of Phase II.

This material is based upon work supported by IFE COLoR under U.S. Department of Energy Grant No. DE-SC0024863, US DOE National Nuclear Security Administration (NNSA) Center of Excellence under Cooperative Agreement No. DE-NA0003869, and by the Department of Energy National Nuclear Security Administration under Award Number DE-NA0004144, the University of Rochester, and the New York State Energy Research and Development Authority. This research used resources of the National Energy Research Scientific Computing Center, a DOE Office of Science User Facility supported by the Office of Science of the U.S. Department of Energy under Contract No. DE-AC02-05CH11231 using NERSC award FES-ERCAP0026741. A. J. thanks useful discussions with B. Afeyan to motivate this study.




\bibliographystyle{jpp}
\bibliography{nlfs}









\end{document}